\documentclass[12pt]{iopart}

\newcommand{\be}{\begin{equation}}
\newcommand{\ee}{\end{equation}}
\newcommand{\bea}{\begin{eqnarray}}
\newcommand{\eea}{\end{eqnarray}}
\newcommand{\beaa}{\begin{eqnarray*}}
\newcommand{\eeaa}{\end{eqnarray*}}
\newcommand{\nn}{\nonumber \\}

\begin{document}

\tolerance=5000

\title{Modified gravity as an alternative for Lambda-CDM cosmology}\footnote{
Based on the invited talk at IRGAC Barcelona conference.}

\author{Shin'ichi Nojiri
\footnote{Electronic address: nojiri@phys.nagoya-u.ac.jp}}
\address{Department of Physics, Nagoya University, Nagoya 464-8602. Japan}

\author{Sergei D. Odintsov
\footnote{Electronic address: odintsov@ieec.uab.es, also at TSPU, Tomsk}}
\address{Instituci\`{o} Catalana de Recerca i Estudis Avan\c{c}ats (ICREA)
and Institut de Ciencies de l'Espai (IEEC-CSIC),
Campus UAB, Facultat de Ciencies, Torre C5-Par-2a pl, E-08193 Bellaterra
(Barcelona), Spain}

\begin{abstract}

The reconstruction scheme is developed for modified $f(R)$ gravity with
realistic  matter (dark matter, baryons, radiation). Two versions of such
theory are constructed: the first one
describes the sequence of radiation and matter domination,
decceleration-acceleration transition  and
acceleration era and the second one is reconstructed from exact
$\Lambda$CDM cosmology. The asymptotic behaviour of first model
at
late times coincides with the theory containing positive and negative
powers of curvature while second model approaches to General Relativity
without  singularity at zero curvature.

\end{abstract}

\maketitle

\section{Introduction}

The modified $f(R)$ gravity (for a review, see \cite{review}) suggests
very interesting gravitational
alternative for dark energy where the current cosmic speed-up is explained
by the presence of some sub-dominant terms (like $1/R$ \cite{c,CDTT} or
$lnR$ \cite{grg}) which may be caused by string/M-theory \cite{no1}.
These terms may become essential at small curvature. There are simple
models of modified $f(R)$ gravity like the one with negative and positive
powers of curvature \cite{NO} which are consistent with late-time
astrophysical constraints \cite{constrain,hall} and solar system tests
\cite{newton}. The accelerating cosmology in such theories has been
studied in refs.\cite{NO,cosmology}.

It became clear recently that modified $f(R)$ gravity may also describe
the sequence of matter dominated, transition from decceleration to
acceleration and acceleration eras \cite{fRany,CNOT}. It has been
developed the reconstruction scheme which permits to reconstruct modified
gravity for any given FRW cosmology \cite{fRany}. In the present
communication we show that such reconstruction scheme may be applied
also in the realistic situation when usual matter is added to modified
$f(R)$ gravity. Specifically, we find two implicitly given versions of
$f(R)$ gravity with matter which may serve as an alternative for
$\Lambda$CDM cosmology. The first model describes the sequence of radiation
dominated, matter dominated and acceleration eras while second model
matches exactly with $\Lambda$CDM cosmology.
Moreover, in the acceleration epoch the asymptotic
behaviour of such theories may be defined: in first case the model
\cite{NO} containing negative power of curvature is recovered while second
model asymptotically approaches to
standard General Relativity.

\section{General formulation of the reconstruction scheme}

In the present section we review the reconstruction scheme
for modified gravity with $f(R)$ action \cite{fRany}, where it has been shown
how any given cosmology may define the implicit form of the function $f$.
The starting action of modified gravity is:
\be
\label{FR1}
S=\int d^4 x \sqrt{-g}f(R)\ .
\ee
The above action is equivalently rewritten as (see, for example,
\cite{capo})
\be
\label{PQR1}
S=\int d^4 x \sqrt{-g} \left\{P(\phi) R + Q(\phi) + {\cal L}_{\rm matter}\right\}\ .
\ee
Here $P$ and $Q$ are proper functions of the scalar field $\phi$ and ${\cal L}_{\rm matter}$ is
the matter Lagrangian density.
Since the scalar field does not have a kinetic term, it may be regarded as an auxiliary field.
In fact, by the variation of $\phi$, it follows $0=P'(\phi)R + Q'(\phi)$,
which may be solved with respect to $\phi$ as $\phi=\phi(R)$.
By substituting the obtained expression of $\phi(R)$ into (\ref{PQR1}),
one comes back to $f(R)$-gravity:
\be
\label{PQR4}
S=\int d^4 x \sqrt{-g} \left\{f(R) + {\cal L}_{\rm matter}\right\}\ , \quad
f(R)\equiv P(\phi(R)) R + Q(\phi(R))\ .
\ee
By the variation of the action (\ref{PQR1}) with respect
to the metric $g_{\mu\nu}$, we obtain
the equations corresponding to standard spatially-flat FRW universe
\bea
\label{PQR6}
0&=&-6 H^2 P(\phi) - Q(\phi) - 6H\frac{dP(\phi(t))}{dt} + \rho \ ,\\
\label{PQR7}
0&=&\left(4\dot H + 6H^2\right)P(\phi) + Q(\phi) + 2\frac{d^2 P(\phi(t))}{dt} + 4H\frac{d P(\phi(t))}{dt} + p\ .
\eea
Simple algebra leads to the following equation\cite{fRany}
\be
\label{PQR7b}
0=2\frac{d^2 P(\phi(t))}{dt^2} - 2 H \frac{dP(\phi(t))}{d\phi} + 4\dot H P(\phi) + p + \rho\ .
\ee
As one can redefine the scalar field $\phi$ freely, we may choose
$\phi=t$.
It is assumed that $\rho$ and $p$ are the sum from the contribution of the matters
with a constant equation of state parameters $w_i$.
Especially, when it is assumed a combination of the radiation and dust, one gets the standard expression
\be
\label{PQR9}
\rho=\rho_{r0} a^{-4} + \rho_{d0} a^{-3}\ ,\quad p=\frac{\rho_{r0}}{3}a^{-4}\ ,
\ee
with constant $\rho_{r0}$ and $\rho_{d0}$. If the scale factor $a$ is given by a proper function $g(t)$
as $a=a_0\e^{g(t)}$ with a constant $a_0$, Eq.(\ref{PQR7}) reduces to the second rank differential
equation (see also \cite{CNOT}):
\bea
\label{PQR11}
0 &=& 2 \frac{d^2 P(\phi)}{d\phi^2} - 2 g'(\phi) \frac{dP(\phi))}{d\phi} + 4g''(\phi) P(\phi) \nn
&& + \sum_i \left(1 + w_i\right) \rho_{i0} a_0^{-3(1+w_i)} \e^{-3(1+w_i)g(\phi)} \ .
\eea
In principle, by solving (\ref{PQR11})  the form of $P(\phi)$ may be found.
Using (\ref{PQR6}) (or equivalently (\ref{PQR7})), the form of $Q(\phi)$
follows as
\be
\label{PQR12}
Q(\phi)=-6 \left(g'(\phi)\right)^2 P(\phi) - 6g'(\phi) \frac{dP(\phi)}{d\phi} \nn
+ \sum_i \rho_{i0} a_0^{-3(1+w_i)}  \e^{-3(1+w_i)g(\phi)} \ .
\ee
Hence, in principle, any given cosmology expressed as $a=a_0\e^{g(t)}$
can be realized as the solution of some specific (reconstructed) $f(R)$-gravity.
Moreover, this reconstruction scheme as is shown in ref.\cite{sami} may be
generalized for other types of modified gravity.

\section{Models of $f(R)$ gravity with transition of matter dominated phase to the acceleration phase}

Let us consider realistic examples where the total action contains also usual matter.
The starting form of $g(\phi)$ is
\be
\label{PQR24}
g(\phi)=h(\phi) \ln \left(\frac{\phi}{\phi_0}\right)\ ,
\ee
with a constant $\phi_0$. It is assumed  that $h(\phi)$ is a slowly changing function of $\phi$.
Due to $\phi=t$, it follows $H\sim h(t)/t$, and the effective EoS (equation of state)
parameter $w_{\rm eff}$ is given by
\be
\label{eff}
w_{\rm eff} \equiv -1 + \frac{2\dot H}{H^2} \sim -1 + \frac{2}{3h(t)}\ .
\ee
Therefore $w_{\rm eff}$ changes slowly with time.

As $h(\phi)$ is assumed to be a a slowly changing function of $\phi=t$,
one can use adiabatic approximation and neglect the derivatives of
$h(\phi)$ like
$\left(h'(\phi)\sim h''(\phi) \sim 0\right)$.
Then Eq.(\ref{PQR11}) has the following form:
\be
\label{PQR25}
0= \frac{d^2 P(\phi)}{d\phi^2} - \frac{h(\phi)}{\phi} \frac{dP(\phi))}{dt} - \frac{2h(\phi)}{\phi^2}  P(\phi) \nn
+ \sum_i \rho_{i0} a_0^{-3(1+w_i)} \e^{-3(1+w_i)g(\phi)} \ .
\ee
The solution for $P(\phi)$ is found to be \cite{fRany}
\be
\label{PQR26}
P(\phi) = p_+ \phi^{n_+(\phi)} + p_- \phi^{n_-(\phi)} + \sum_i p_i(\phi) \phi^{-3(1+w_i)h(\phi) + 2} \ .
\ee
Here $p_\pm$ are arbitrary constants and
\bea
\label{PQR27}
n_\pm (\phi) &\equiv& \frac{h(\phi) - 1 \pm \sqrt{h(\phi)^2 + 6h(\phi) + 1}}{2}\ ,\nn
p_i (\phi) &\equiv& - \left\{(1+w_i)\rho_{i0} a_0^{-3(1+w_i)} \phi_0^{3(1+w_i)h(\phi)}\right\} \nn
&& \times \left\{6(1+w_i)(4+3w_i) h(\phi)^2 - 2 \left(13 + 9w_i\right)h(\phi) + 4\right\}^{-1}\ .
\eea
Especially for the radiation and dust, one has
\bea
\label{PP2}
p_{\rm rad}r (\phi) & \equiv & - \frac{4\rho_{r0}\phi_0^{4h(\phi)} }{3a_0^4
\left( 40 h(\phi)^2 - 32 h(\phi) + 4\right)}\ ,\nn
p_{\rm dust} (\phi) & \equiv & - \frac{\rho_{d0}\phi_0^{3h(\phi)} }{a_0^3
\left( 24 h(\phi)^2 - 26 h(\phi) + 4\right)}\ .
\eea
The form of $Q(\phi)$ is found to be
\bea
\label{PQR28}
Q(\phi) &=& - 6h(\phi)p_+ \left(h(\phi) + n_+(\phi) \right) \phi^{n_+ (\phi) - 2} \nn
&& - 6h(\phi)p_- \left(h(\phi) + n_-(\phi) \right) \phi^{n_- (\phi) - 2} \nn
&& + \sum_i\left\{ - 6h(\phi) \left( -(2+3w_i)h(\phi) + 2\right)p_i (\phi) \right. \nn
&& \left. + p_{i0} a_0^{-3(1+w_i)}\phi_0^{3(1+w_i)h(\phi)}\right\} \phi^{- 3(1+w_i)h(\phi)} \ .
\eea
Eq.(\ref{PQR24}) tells that
\be
\label{PQE56}
R\sim \frac{6\left( -h(t) + 2h(t)^2\right)}{t^2}\ .
\ee
Let assume $\lim_{\phi\to 0} h(\phi) = h_i$ and $\lim_{\phi\to \infty} h(\phi) = h_f$.
Then if $0<h_i<1$, the early universe is in decceleration phase and
if $h_f>1$, the late universe is in acceleration phase.
We may consider the case $h(\phi)\sim h_m$ is almost constant when $\phi\sim t_m$
$\left(0\ll t_m \ll +\infty\right)$.
If $h_1$, $h_f>1$ and $0<h_m<1$,  the early universe is also accelerating, which
could describe the inflation. After that the universe becomes
deccelerating,
which corresponds to matter-dominated phase with $h(\phi)\sim 2/3$ there.
Furthermore, after that, the universe enters the acceleration epoch.
Hence, the unification of the inflation, matter domination and
late-time acceleration is possible in the theory under consideration.

As an extension of the above model \cite{fRany}, we consider the inclusion
of the
radiation, baryons, and dark matter:
\be
\label{PQ1}
h(\phi) = \frac{h_i + h_f q \phi^2}{1 + q \phi^2}\ ,
\ee
with constants $h_i$, $h_f$, and $q$. When $\phi\to 0$, $h(\phi)\to h_i$ and
when $\phi\to \infty$, $h(\phi)\to h_f$. If $q$ is small enough, $h(\phi)$ can be a slowly
varying function of $\phi$.
By using the expression of (\ref{PQE56}), we find\cite{fRany}
\bea
\label{PQ2}
&& \phi^2=\Phi_0(R)\ ,\quad \Phi_\pm (R)\ ,\nn
&& \Phi_0 \equiv \alpha_+^{1/3} + \alpha_-^{1/3}\ ,\quad
\Phi_\pm \equiv \alpha_\pm^{1/3}\e^{2\pi i/3} + \alpha_\mp^{1/3}\e^{-2\pi i/3}\ ,\nn
&& \alpha_\pm \equiv \frac{-\beta_0 \pm \sqrt{\beta_0^2 - \frac{4\beta_1^3}{27}}}{2}\ ,\nn
&& \beta_0 \equiv \frac{2\left(2R + 6h_f q - 12 h_f^2 q \right)^3}{27 q^3 R^3} \nn
&& \qquad - \frac{\left(2R + 6h_f q - 12 h_f^2 q \right)\left(R + 6h_i q + 6 h_f q - 4h_i h_f q\right)}{3qR} \nn
&& \qquad + 6h_i - 12 h_i^2\ ,\nn
&& \beta_1 \equiv - \frac{\left(2R + 6h_f q - 12 h_f^2 q \right)^2}{3 q^2 R^2}
 - \frac{R + 6h_i q + 6 h_f q - 4h_i h_f q}{q^2 R}\ .
\eea
There are three branches $\Phi_0$ and $\Phi_\pm$ in (\ref{PQ2}). Eqs.(\ref{PQE56}) and (\ref{PQ1}) show
that when the curvature is small ($\phi=t$ is large), we find $R\sim 6\left( - h_f + 2 h_f^2 \right)/\phi^2$
and when the curvature is large ($\phi=t$ is small), $R\sim 6\left( - h_i + 2 h_i^2 \right)/\phi^2$.
This asymptotic behaviour indicates that we should choose $\Phi_0$ in (\ref{PQ2}).
Then explicit form of $f(R)$ could be given by using the expressions of $P(\phi)$  (\ref{PQR26}) and
$Q(\phi)$  (\ref{PQR28}) as
\be
\label{PQ3}
f(R)=P\left(\sqrt{\Phi_0(R)}\right) R + Q\left(\sqrt{\Phi_0(R)}\right)\ .
\ee

One may check the asymptotic behavior of $f(R)$  (\ref{PQ3}) (for some
parameters choice) in
the acceleration era coincides with the theory proposed in \cite{NO}.
We now consider the case that, besides dust, which coul be dark matter and baryons, there is a radiation.
In this case, $P(\phi)$ is given by
\bea
\label{IR1}
P(\phi) &=& p_+ \phi^{n_+(\phi)} + \left(p_{\rm dark}(\phi) + p_{\rm baryon}(\phi)\right)\phi^{-3h(\phi) + 2}
+ p_{\rm rad}(\phi) \e^{-4h(\phi) + 2}\ .
\eea
Here $p_{\rm rad}(\phi)$ is given by (\ref{PP2}) and $p_{\rm dark}(\phi) + p_{\rm baryon}(\phi)$ are
\be
\label{IR2}
p_{\rm dark} (\phi)  + p_{\rm baryon}(\phi) \equiv
 - \frac{\left(\rho_{{\rm dark}0} + \rho_{{\rm baryon}0}\right)
\phi_0^{3h(\phi)} }{a_0^3\left( 24 h(\phi)^2 - 26 h(\phi) + 4\right)}\ .
\ee
We now find $n_+ >  - 3h(\phi) + 2 > - 4h(\phi) + 2 >0$, in (\ref{IR1}).
Here $n_+=\left(h(\phi) - 1 \pm \sqrt{h(\phi)^2 + 6h(\phi) + 1}\right)/2$ is defined in (\ref{PQR27}).
Then when $\phi$ is large, the first term in (\ref{IR1}) dominates and
when $\phi$ is small, the last term dominates.
When $\phi$ is large, curvature is small and  $\phi^2\sim 6\left( - h_f + 2 h_f \right)/R$
and $h(\phi)\to h(\infty)=h_f$.
Hence, $f(R)\sim R^{-\left(h(\phi) - 5 + \sqrt{h_f^2 + 6h_f + 1}\right)/4
}$.
Especially when $h\gg 1$, it becomes $f(R)\sim R^{-h_f/2}$.
Therefore there appears the negative power of $R$ predicted by the
presence of matter-dominated stage.
As $H\sim h_f/t$, if $h_f>1$, the universe is in acceleration phase.

On the other hand, when curvature is large, we find
$\phi^2\sim 6\left( - h_i + 2 h_i \right)/R$ and $h(\phi)\to h(0)=h_i$.
If the universe era corresponds to radiation dominated phase ($h_i=1/2$),
$P(\phi)$ becomes a constant and therefore $f(R)\sim R$,
which reproduces the Einstein gravity.

Thus, in the above model,the  radiation/matter dominated phase evolves
into acceleration phase and $f(R)$ behaves as
$f(R)\sim R$ initially while $f(R)\sim R^{-\left(h(\phi) - 5 + \sqrt{h_f^2 + 6h_f + 1}\right)/4 }$ at late time.
For the matter dominated phase, we have $h=2/3$. Since $h_i=1/2<2/3<h_f$ and $h(\phi=t)$ is a slowly increasing
function of $\phi=t$, there should always be a matter dominated phase.
Therefore in the model (\ref{PQ1}) with $h_i=1/2$, the universe is first
in radiation dominated phase. Subsequently, the
universe evolves to the matter dominated phase, and finally to accelerated
phase which is consistent with $\Lambda$CDM cosmology.

Thus, we presented the example of implicitly given $f(R)$ gravity which
describes the radiation dominated era, the
matter dominated stage, transition from decceleration to acceleration and
acceleration epoch (where it may include the negative powers of $R$). This
model seems to be quite reasonable alternative for the standard
$\Lambda$CDM
cosmology.

\section{Model reproducing $\Lambda$CDM-type cosmology}

Let us investigate if $\Lambda$CDM-type cosmology could be reconstructed
exactly by $f(R)$-gravity in the present formulation when we include dust,
which could be a sum of the baryon and dark matter, and radiation.

In the Einstein gravity, when there is a matter with the EOS parameter $w$
and cosmological constant, the FRW equation has the following form:
\be
\label{LCDM1}
\frac{3}{\kappa^2}H^2 = \rho_0 a^{-3(1+w)} + \frac{3}{\kappa^2 l^2}\ .
\ee
Here $l$ is the length parameter.
The solution of (\ref{LCDM1}) is given by
\be
\label{LCDM2}
a=a_0\e^{g(t)}\ ,\quad
g(t)=\frac{2}{3(1+w)}\ln \left(\alpha \sinh
\left(\frac{3(1+w)}{2l}\left(t - t_0 \right)\right)\right)\ .
\ee
Here $t_0$ is a constant of the integration and
$\alpha^2\equiv (1/3)\kappa^2 l^2 \rho_0 a_0^{-3(1+w)}$.
Let us show how  is possible to reconstruct $f(R)$-gravity reproducing (\ref{LCDM2}).
When matter is included, Eq.(\ref{PQR11})  has the following form:
\bea
\label{LCDM4}
0 &=& 2\frac{d^2 P(\phi)}{d\phi^2} - \frac{2}{l}\coth \left(\frac{3(1+w)}{2l}\left(t\phi- t_0 \right)\right) \frac{dP(\phi)}{d\phi} \nn
&& - \frac{6(1+w)}{l^2} \sinh^{-2} \left(\frac{3(1+w)}{2l}\left(\phi - t_0 \right)\right) P(\phi) \nn
&& + \frac{4}{3}\rho_{r0} a_0^{-4} \left(\alpha \sinh \left(\frac{3(1+w)}{2l}\left(\phi - t_0 \right)\right)\right)^{-8/3(1+w)} \nn
&& + \rho_{d0}a_0^{-3}\left(\alpha \sinh \left(\frac{3(1+w)}{2l}\left(\phi - t_0 \right)\right)\right)^{-2/(1+w)}\ .
\eea
Since this equation is the linear inhomogeneous equation,
 the general solution is given by the sum of the
special solution and the general solution which corresponds to  the homogeneous equation.
For the case without matter, by changing the variable from $\phi$ to $z$ as follows,
\be
\label{LCDM5}
z\equiv - \sinh^{-2} \left(\frac{3(1+w)}{2l}\left(t - t_0 \right)\right) \ ,
\ee
Eq.(\ref{LCDM4}) without matter can be rewritten in the form of Gauss's hypergeometric differential equation:
\bea
\label{LCDM6}
&& 0=z(1-z)\frac{d^2 P}{dz^2} + \left[\tilde\gamma - \left(\tilde\alpha + \tilde \beta + 1\right)z\right] \frac{dP}{dz}
 - \tilde\alpha \tilde\beta P\ , \nn
&& \tilde\gamma \equiv 4 + \frac{1}{3(1+w)},\
\tilde\alpha + \tilde\beta + 1 \equiv 6 + \frac{1}{3(1+w)},\
\tilde\alpha \tilde\beta \equiv - \frac{1}{3(1+w)},
\eea
whose solution is given by Gauss's hypergeometric function:
\be
\label{LCDM7}
P= P_0 F(\tilde\alpha,\tilde\beta,\tilde\gamma;z)
\equiv P_0 \frac{\Gamma(\tilde\gamma)}{\Gamma(\tilde\alpha) \Gamma(\tilde\beta)}
\sum_{n=0}^\infty \frac{\Gamma(\tilde\alpha + n) \Gamma(\beta + n)}{\Gamma(\tilde\gamma + n)}
\frac{z^n}{n!}\ .
\ee
Here $\Gamma$ is the $\Gamma$-function. There is one more linearly independent solution like
$(1-z)^{\tilde\gamma - \tilde\alpha - \tilde\beta}
F(\tilde\gamma - \tilde\alpha, \tilde\gamma - \tilde\beta, \tilde\gamma;z)$
but we drop it here, for simplicity.
Using (\ref{PQR12}), one finds the form of $Q(\phi)$:
\bea
\label{LCDM8}
Q &=& - \frac{6(1-z)P_0}{l^2}F( \tilde\alpha,\tilde\beta,\tilde\gamma;z) \nn
&& - \frac{3(1+w) z(1-z)P_0}{l^2(13 + 12w)}
F(\tilde\alpha+1,\tilde\beta+1,\tilde\gamma+1;z)\ .
\eea
 From (\ref{LCDM5}), it follows $z\to 0$ when $t=\phi\to + \infty$. Then
in the limit, one arrives at $P(\phi)R + Q(\phi) \to P_0 R - 6P_0/l^2$.
Identifying $P_0=1/2\kappa^2$ and $\Lambda = 6/l^2$,
the Einstein theory with cosmological constant $\Lambda$ can be reproduced.
The action is not singular even in the limit of $t\to \infty$.
Therefore even without cosmological constant nor cold dark matter, the cosmology of $\Lambda$CDM model
could be reproduced by $f(R)$-gravity (for a different treatment,
see \cite{dobado}).

We now investigate the special solution of (\ref{LCDM4}).
By changing the variable as in (\ref{LCDM5}), the
inhomogeneous differential equation looks as:
\bea
\label{LCDMa1}
&& 0=z(1-z)\frac{d^2 P}{dz^2} + \left[\tilde\gamma - \left(\tilde\alpha + \tilde \beta + 1\right)z\right] \frac{dP}{dz}
 - \tilde\alpha \tilde\beta P \nn
&& \qquad + \eta\left(-z\right)^{-2(1+3w)/3(1+w)} + \xi\left(-z\right)^{-\frac{1+2w}{1+w}}\ , \nn
&& \eta \equiv \frac{4l^2}{27(1+w)}\rho_{r0}a_0^{-4}\alpha^{-8/3(1+w)}\ ,\quad
\zeta \equiv \frac{l^2}{3(1+w)}\rho_{d0}a_0^{-3}\alpha^{-2/(1+w)}\ .
\eea
It is not trivial to find the solution of (\ref{LCDMa1}).
Let us  consider the case that $w=0$ and $z\to -\infty$, that is,
$t\to t_0$. In the limit,
Eq.(\ref{LCDMa1}) reduces to
\be
\label{LCDMa2}
0= - z^2\frac{d^2 P}{dz^2} + \tilde\gamma z \frac{dP}{dz} - \tilde\alpha \tilde\beta P + \eta\left(-z\right)^{-2/3} \ ,
\ee
whose special solution is given by
\be
\label{LCDMa3}
P=P_0(-z)^{-2/3}\ ,\quad P_0 = \frac{\eta}{\frac{10}{9} - \frac{2\left(\tilde\alpha + \tilde\beta + 1\right)}{3}
+ \tilde\alpha\tilde\beta} = - \frac{9\eta}{25}\ .
\ee
In principle, there could be found other special solution of Eq.
(\ref{LCDMa1}).
This proves that  even in the presence of matter, the standard
$\Lambda$CDM cosmology
could be reproduced by $f(R)$-gravity exactly.

Thus, we presented two versions of modified $f(R)$ gravity with matter.
The first
version describes the sequence of radiation dominated, matter dominated,
transition from decceleration to acceleration and acceleration eras
(compare with scalar-tensor gravity with the same emerging cosmology
\cite{stefancic}). In the aceleration era the action may asymptotically
approach to the action with negative and positive powers of $R$ proposed
in \cite{NO}. Moreover, for some choice of parameters it reproduces the
$\Lambda$CDM cosmology at late times. The second model may reproduce
$\Lambda$CDM
cosmology exactly. Using the number of free parameters of the models one
can
expect that they may be in good correspondence with observatonal data as
they are with three years WMAP data. Nevertheless, the precise fitting of
the proposed $f(R)$ gravity against existing/coming observational data
should be done.

Acknowledgements.
We thank J. Sola for the invitation to give the talk at IRGAC conference.
The work by S.N. was supported in part by Monbusho grant  no.18549001
(Japan) and
21st
Century COE program of Nagoya Univ. provided by JSPS (15COEEG01) and that
by S.D.O. by the project FIS2005-01181 (MEC, Spain) and by RFBR grant
06-01-00609 (Russia).


\section*{References}

\begin{thebibliography}{99}

\bibitem{review}
S.~Nojiri and S.~D.~Odintsov,
arXiv:hep-th/0601213.

\bibitem{c}
S.~Capozziello,
Int.\ J.\ Mod.\ Phys.\ D {\bf 11} (2002) 483;
S.~Capozziello, S.~Carloni and A.~Troisi,
arXiv:astro-ph/0303041.

\bibitem{CDTT}
S.~M.~Carroll, V.~Duvvuri, M.~Trodden and S.~Turner,
Phys.\ Rev.\ D {\bf 70} (2004) 043528.

\bibitem{grg}
S.~Nojiri and S.~D.~Odintsov,
GRG {\bf 36} (2004) 1765
[arXiv:hep-th/0308176].

\bibitem{no1}
S.~Nojiri and S.~D.~Odintsov,
Phys.\ Lett.\ B {\bf 576} (2003) 5
[arXiv:hep-th/0307071].

\bibitem{NO}
S.~Nojiri and S.~D.~Odintsov,
Phys.\ Rev.\ D {\bf 68}, (2003) 123512
[arXiv:hep-th/0307288].

\bibitem{constrain}
S.~Capozziello, V.~F.~Cardone and A.~Troisi,
Phys.\ Rev.\ D {\bf 71} (2005) 043503;
S.~Capozziello, V.~F.~Cardone and M.~Francaviglia,
GRG {\bf 38} (2006) 711;
M.~Amarzguioui, O.~Elgaroy, D.~F.~Mota and T.~Multamaki,
arXiv:astro-ph/0510519;
O.~Mena, J.~Santiago and J.~Weller,
Phys.\ Rev.\ Lett.\ {\bf 96} (2006) 041103;
T.~Koivisto and H.~Kurki-Suonio,
Class.\ Quant.\ Grav.\ {\bf 23} (2006) 2355;
S.~Capozziello, V.~F.~Cardone, E.~Elizalde, S.~Nojiri and S.~D.~Odintsov,
Phys.\ Rev.\ D {\bf 73} (2006) 043512,
[arXiv:astro-ph/0508350];
A.~Borowiec, W.~Godlowski and M.~Szydlowski,
arXiv: astro-ph/0607639;
S.~Bludman,
arXiv:astro-ph/0605198;
S.~Capozziello, A.~Stabile and A.~Troisi,
arXiv:gr-qc/0603071;
S. Bergliaffa, arXiv: gr-qc/0608072;
D.~Huterer and E.~Linder,
arXiv: astro-ph/0608681.

\bibitem{hall}
A.~W.~Brookfield, C.~van~de~Bruck and L.~Hall,
hep-th/0608015.

\bibitem{newton}
G.~Allemandi, M.~Francaviglia, M.~Ruggiero and A.~Tartaglia,
arXiv:gr-qc/0506123;
X.~Meng and P.~Wang,
GRG {\bf 36} (2004) 1947;
A.~Domingues and D.~Barraco,
Phys.\ Rev.\ D {\bf 70} (2004) 043505;
T.~Koivisto,
arXiv:gr-qc/0505128;
T.~Clifton and J.~Barrow,
Phys.\ Rev.\ D {\bf 72} (2005) 103005;
J.~Cembranos,
Phys.\ Rev.\ D {\bf 73} (2006) 064029;
T.~Sotiriou,
arXiv:gr-qc/0507027;
I.~Navarro and K.~Van~Acoleyen,
Phys.\ Lett.\ B {\bf 622} (2005) 1;
C.~Shao, R.~Cai, B.~Wang and R.~Su,
arXiv:gr-qc/0511034;
M.~E.~Soussa and R.~P.~Woodard,
GRG {\bf 36} (2004) 855;
S.~Capozziello and A.~Troisi,
Phys.\ Rev.\ D {\bf 72} (2005) 044022;
K.~Atazadeh and H.~Sepangi,
arXiv:gr-qc/0602028;
R.~Woodard,
arXiv:astro-ph/0601672;
A.~D.~Dolgov and M.~Kawasaki,
Phys.\ Lett.\ B {\bf 573} (2003) 1;
V.~Faraoni,
arXiv:gr-qc/0607016.

\bibitem{cosmology}
S.~Nojiri and S.~D.~Odintsov,
Mod.\ Phys.\ Lett.\ A {\bf 19} (2004) 627
[arXiv:hep-th/0310045];
G.~Allemandi, A.~Borowiec and M.~Francaviglia,
Phys.\ Rev.\ D {\bf 70} (2004) 043524;
X.~Meng and P.~Wang,
Class.\ Quant.\ Grav.\ {\bf 22} (2005) 23;
Class.\ Quant.\ Grav.\ {\bf 21} (2004) 951;
M.~C.~B.~Abdalla, S.~Nojiri and S.~D.~Odintsov,
Class.\ Quant.\ Grav.\ {\bf 22} (2005) L35
[arXiv:hep-th/0409177];
G.~Allemandi, A.~Borowiec, M.~Francaviglia and S.~D.~Odintsov,
Phys.\ Rev.\ D {\bf 72} (2005) 063505
[arXiv:gr-qc/0504057];
D.~Easson,
Int.\ J.\ Mod.\ Phys.\ A {\bf 19} (2004) 5343;
G.~Cognola, E.~Elizalde, S.~Nojiri, S.~D.~Odintsov and S.~Zerbini,
JCAP {\bf 0502} (2005) 010
[arXiv:hep-th/0501096];
T.~Multamaki and I.~Vilja,
Phys.\ Rev.\ D {\bf 73} (2006) 024018;
N.~Poplawski,
arXiv:gr-qc/0510007;
V.~Faraoni,
Phys.\ Rev.\ D{\bf 72} (2005) 124005;
I.~Brevik,
arXiv:gr-qc/0603025; gr-qc/0601100;
T.~Sotiriou,
Phys.\ Rev.\ D {\bf 73} (2006) 063515,
arXiv:gr-qc/0509029;
S.~K.~Srivastava,
arXiv:hep-th/0605010;
D.~Samart,
arXiv:astro-ph/0606612.

\bibitem{fRany}
S.~Nojiri and S.~D.~Odintsov,
arXiv:hep-th/0608008.

\bibitem{CNOT}
S.~Capozziello, S.~Nojiri, S.~D.~Odintsov and A.~Troisi,
arXiv:astro-ph/0604431.

\bibitem{capo}
S.~Capozziello, S.~Nojiri and S.~D.~Odintsov,
Phys.\ Lett.\ B {\bf 634} (2006) 93
[arXiv:hep-th/0512118].

\bibitem{sami}
S. Nojiri, S.D. Odintsov and M. Sami, hep-th/0605039,
Phys.\ Rev.\ D {\bf 74} (2006) 046004.

\bibitem{dobado}
A.~Cruz-Dombriz and A.~Dobado,
Phys.\ Rev.\ D {\bf 74} (2006) 087501.

\bibitem{stefancic}
S.~Nojiri, S.~D.~Odintsov and H.~Stefancic,
arXiv:hep-th/0608168.

\end{thebibliography}
\end{document}